\documentclass[10pt,twocolumn,conference]{IEEEtran}

\IEEEoverridecommandlockouts

\usepackage{graphicx}
\usepackage{amsmath}
\usepackage{amssymb}
\usepackage[caption=false]{subfig}
\usepackage[noadjust]{cite}
\usepackage{float}
\usepackage{algorithm}
\usepackage{bibentry}
\usepackage{balance}
\usepackage{algorithm}
\usepackage{algorithmicx}
\usepackage{algpseudocode}
\usepackage{xcolor}
\usepackage{subfig}

\usepackage{url}

\graphicspath{{img/}}

\begin{document}

\bstctlcite{IEEEexample:BSTcontrol}

\title{SDR-Based 5G NR C-Band I/Q Monitoring and Surveillance in Urban Area Using a Helikite\thanks{This research is supported in part by the NSF award CNS-1939334.}}

\author{Sung Joon Maeng$^*$, Ozgur Ozdemir$^*$, \.{I}smail G\"{u}ven\c{c}$^*$, Mihail L. Sichitiu$^*$,\\
Magreth Mushi$^*$, Rudra Dutta$^*$, and Monisha Ghosh$^\dagger$\\
$^*$Department of Electrical and Computer Engineering, North Carolina State University, Raleigh, NC\\
$^\dagger$Department of Electrical Engineering, University of Notre Dame\\
{\tt \{smaeng,oozdemi,iguvenc,mlsichit,mjmushi,rdutta\}@ncsu.edu mghosh3@nd.edu} }

\maketitle

\begin{abstract}
In this paper, we report experimental results in collectting and processing 5G NR I/Q samples in the 3.7~GHz C-band by using software-defined radio (SDR)-mounted helikite. 
We use MATLAB's 5G toolbox to post-process the collected data, to obtain the synchronization signal block (SSB) from the I/Q samples and then go through the cell search, synchronization procedures, and reference signal received power (RSRP) and and reference signal received quality (RSRQ) calculation. We plot these performance metrics for various physical cell identities as a function of the helikite's altitude.  Furthermore, building on our experience with the collected and post-processed data, we discuss potential  vulnerabilities of  5G NR systems to surveillance, jamming attacks, and post quantum era attacks.
\end{abstract}


\section{Introduction} \label{sec:intro}
Recently, unmanned aerial vehicles (UAVs) have been rapidly gaining attention due to their various potential applications. UAVs can immediately fly into hazardous areas for search and rescue missions, and can be used to transport medical supplies for public safety~\cite{UNICEF}. Base station (BS) mounted on UAVs can provide improved wireless coverage by reducing coverage holes and can add capacity to support larger number of users~\cite{chowdhury20203}. UAVs equipped with 5G capabilities can also be used for data collection and edge computation for massive machine type communications (mMTC)~\cite{zhu2018energy,sabuj2020cognitive}  in smart cities~\cite{horsmanheimo20225g}, smart agriculture~\cite{valecce2019interplay}, and industrial Internet of Things (IIoT) settings~\cite{liang2021intelligent,al20205g}.
 To realize these future applications, it is critical to have wireless connectivity with UAVs in beyond-visual-line-of-sight (BVLOS) scenarios, which can be provided by cellular-connected UAVs (C-UAVs).

Recent studies have examined C-UAVs connectivity and coverage using the widely used 4G long-term evolution (LTE) wireless standard. In \cite{8746290,8108204,8369158}, the use of LTE for serving UAVs has been investigated and the air-to-ground radio propagation has been explored by flying a smartphone-mounted UAV. However, results from these works are limited to a set of key performance indicators (KPIs) available to commercial smartphone software. In \cite{maeng2022aeriq}, air-to-ground channel propagation is studied by LTE eNB and a drone equipped with an SDR and a GPS receiver. The collected raw I/Q sample dataset is post-processed and analyzed using MATLAB's LTE Toolbox to understand multiple aspects of channel propagation and receiver algorithms such as synchronization, channel estimation, and extraction of reference signal received power (RSRP), spatial correlation, coherence time, and coherence bandwidth.   

\begin{figure}[t]
	\centering
        \subfloat[Helikite flying over the main campus of NC State University, Raleigh, NC, with downtown Raleigh in the background.]{\includegraphics[width=0.4\textwidth]{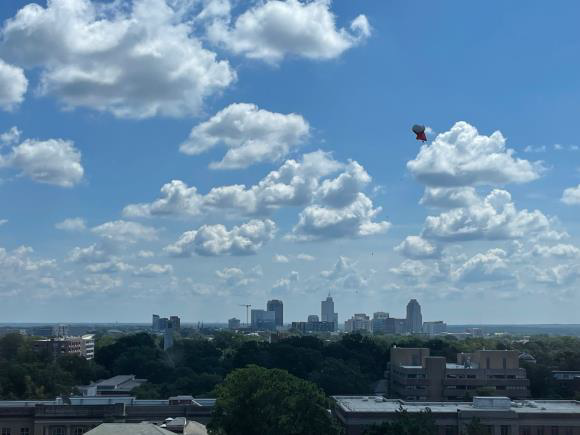}\label{fig:photo_packa}}\vspace{-3mm}\\
        \subfloat[Altitdue of the AERPAW helikite during the experiment, which is obtained by GPS logs. The helikite flies up and down four times.]{\includegraphics[width=0.49\textwidth]{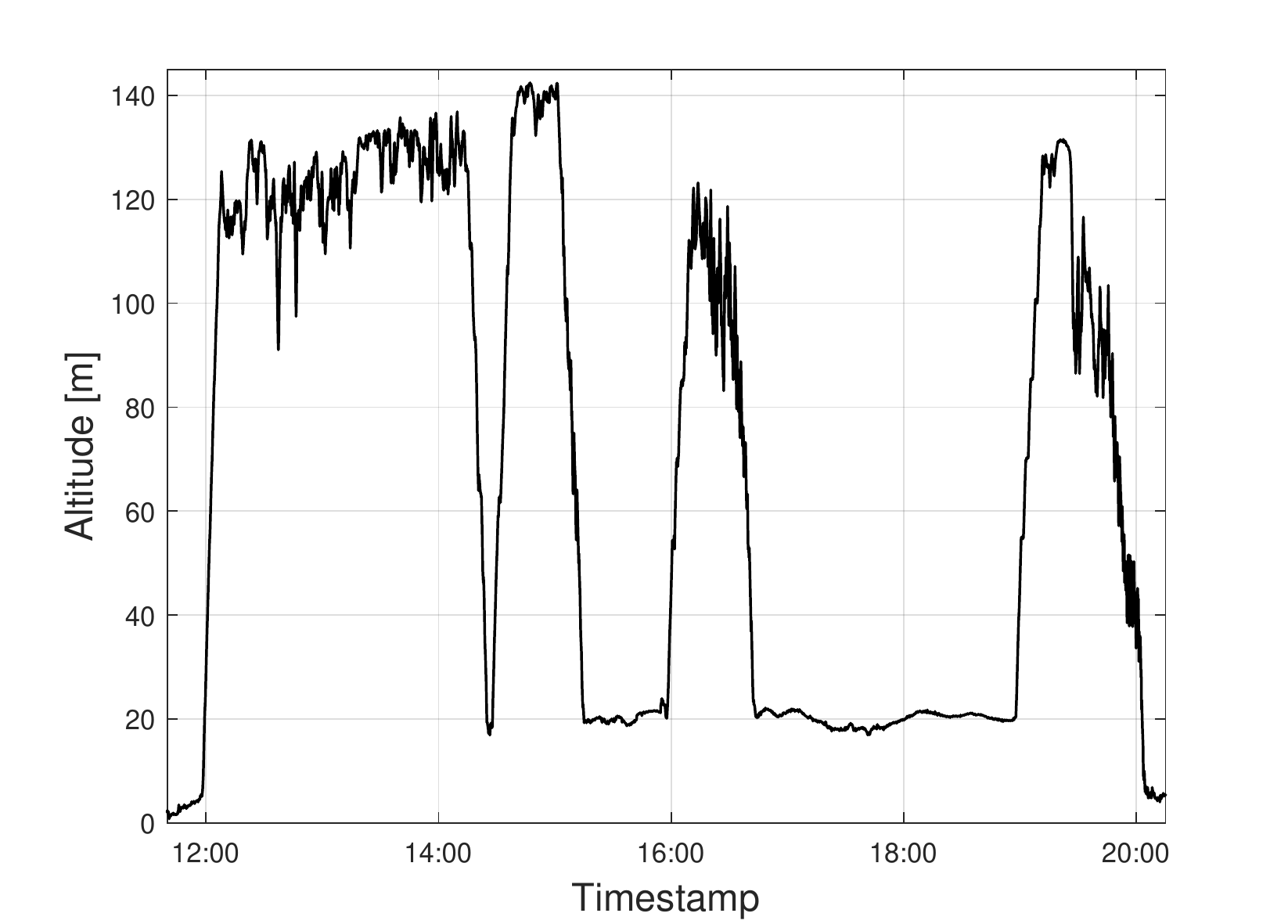}\label{fig:altitude}}
	\caption{Helikite experiment during NC State’s Packapalooza festival close to downtown Raleigh.}
\end{figure}

5G New Radio (NR) is an advanced cellular wireless communication standard taking over LTE. It has also been studied to some extent in the literature for providing wireless coverage to UAVs~\cite{horsmanheimo20225g,heikkila2022latency}. Although 5G NR is more secure than LTE, 5G NR is still vulnerable to cyber security attacks. In \cite{lichtman20185g,arjoune2020smart,girke2019towards}, the vulnerability of 5G NR against jamming, spoofing, and sniffing attacks has been studied, and mitigation techniques are recommended. In particular, synchronization signals (SS), physical broadcast channel (PBCH), and the physical broadcast channel demodulation reference signal symbols (PBCH-DMRS) in a synchronization signal block (SSB) are easy targets from an adversary in terms of attack efficiency and complexity~\cite{lichtman20185g}. Moreover, the arrival of quantum computing is expected to break most of the public-key security schemes due to higher computation capabilities~\cite{garcia2022disruptive}.

\begin{figure*}[t!]
	\centering
        \subfloat[Center frequency: 3755.68~MHz]{\includegraphics[width=0.48\textwidth]{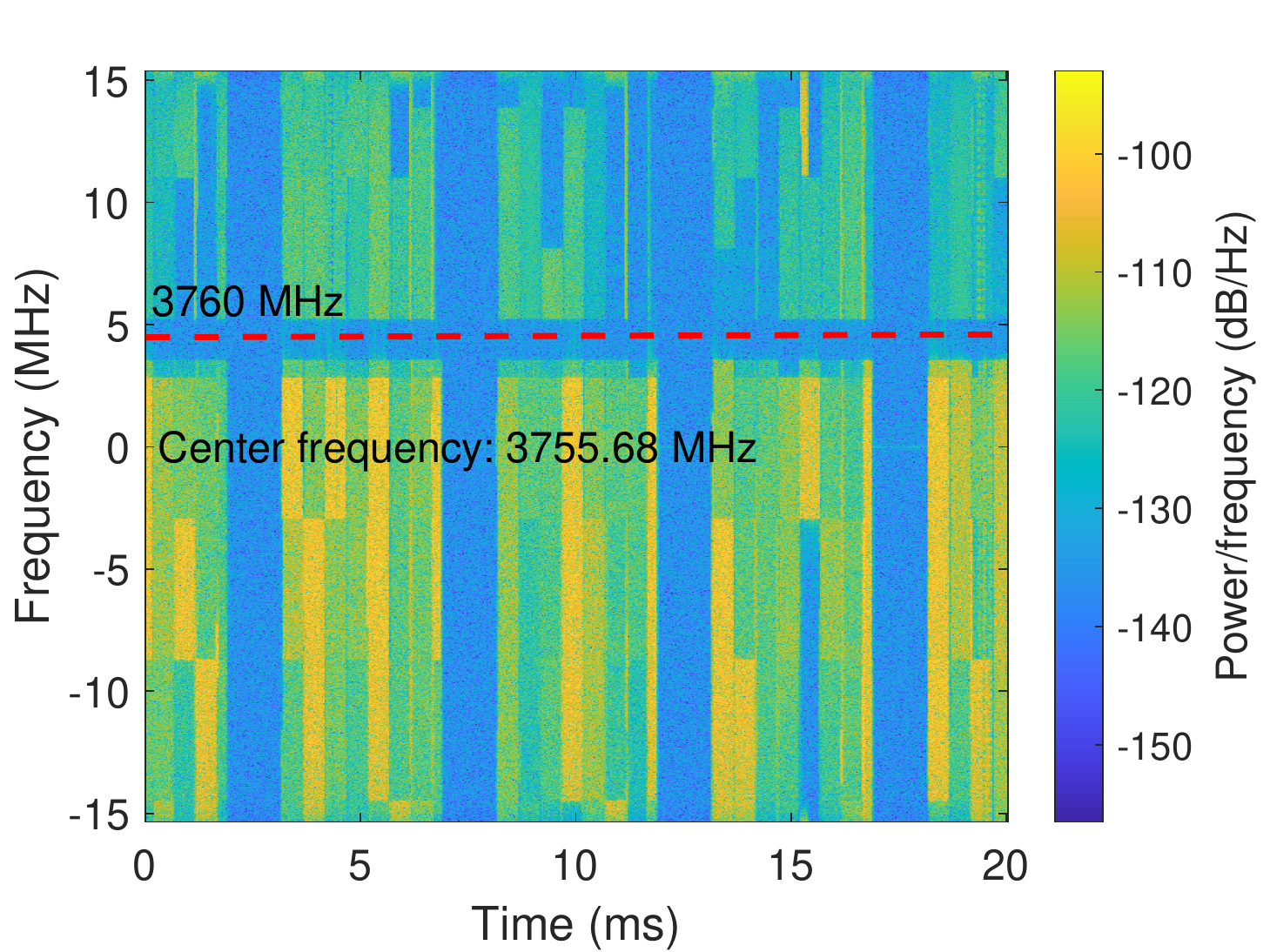}}
        \subfloat[Center frequency: 3728.80~MHz]{\includegraphics[width=0.48\textwidth]{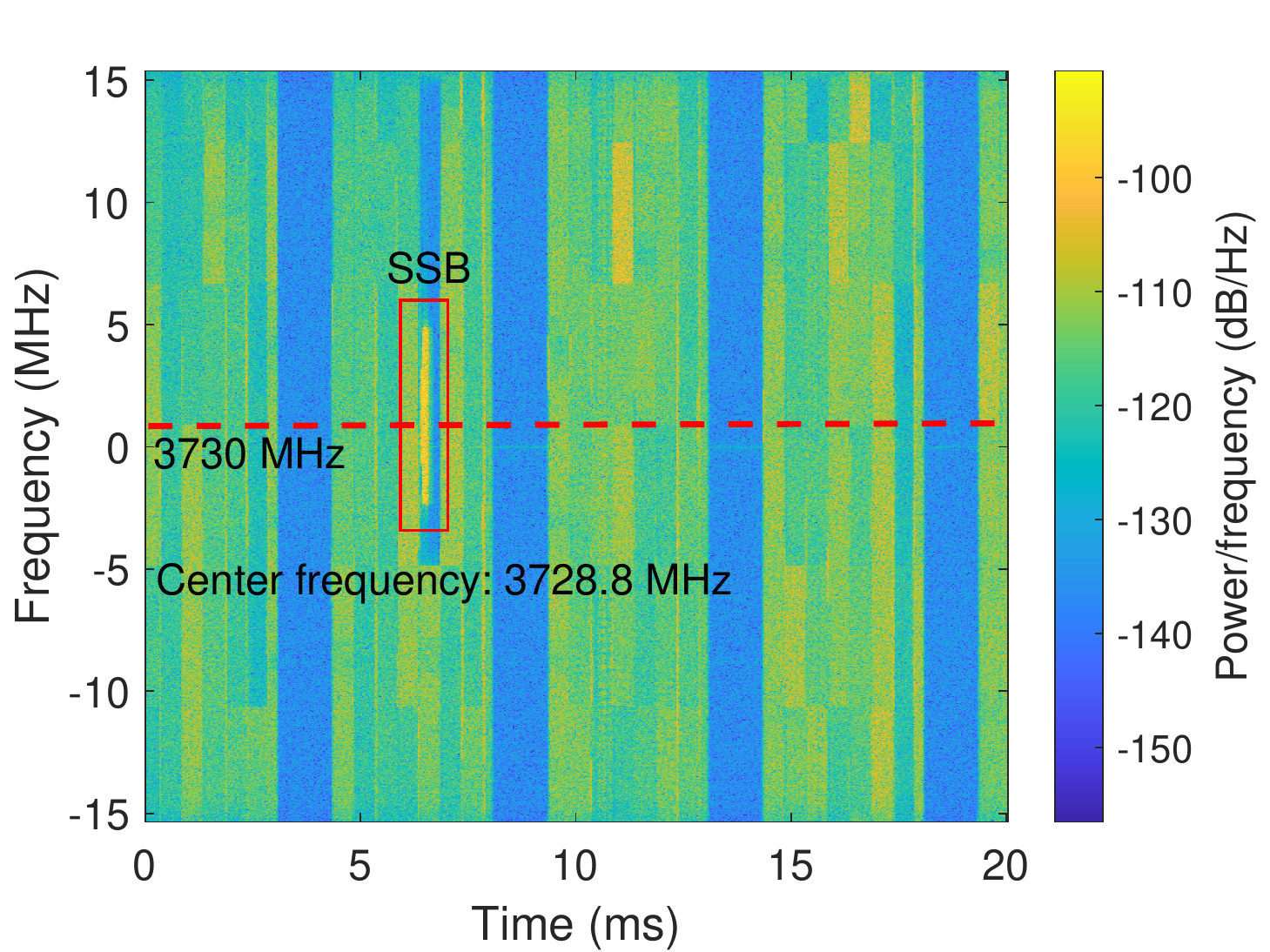}}\\

        \subfloat[Center frequency: 3708.64~MHz]{\includegraphics[width=0.48\textwidth]{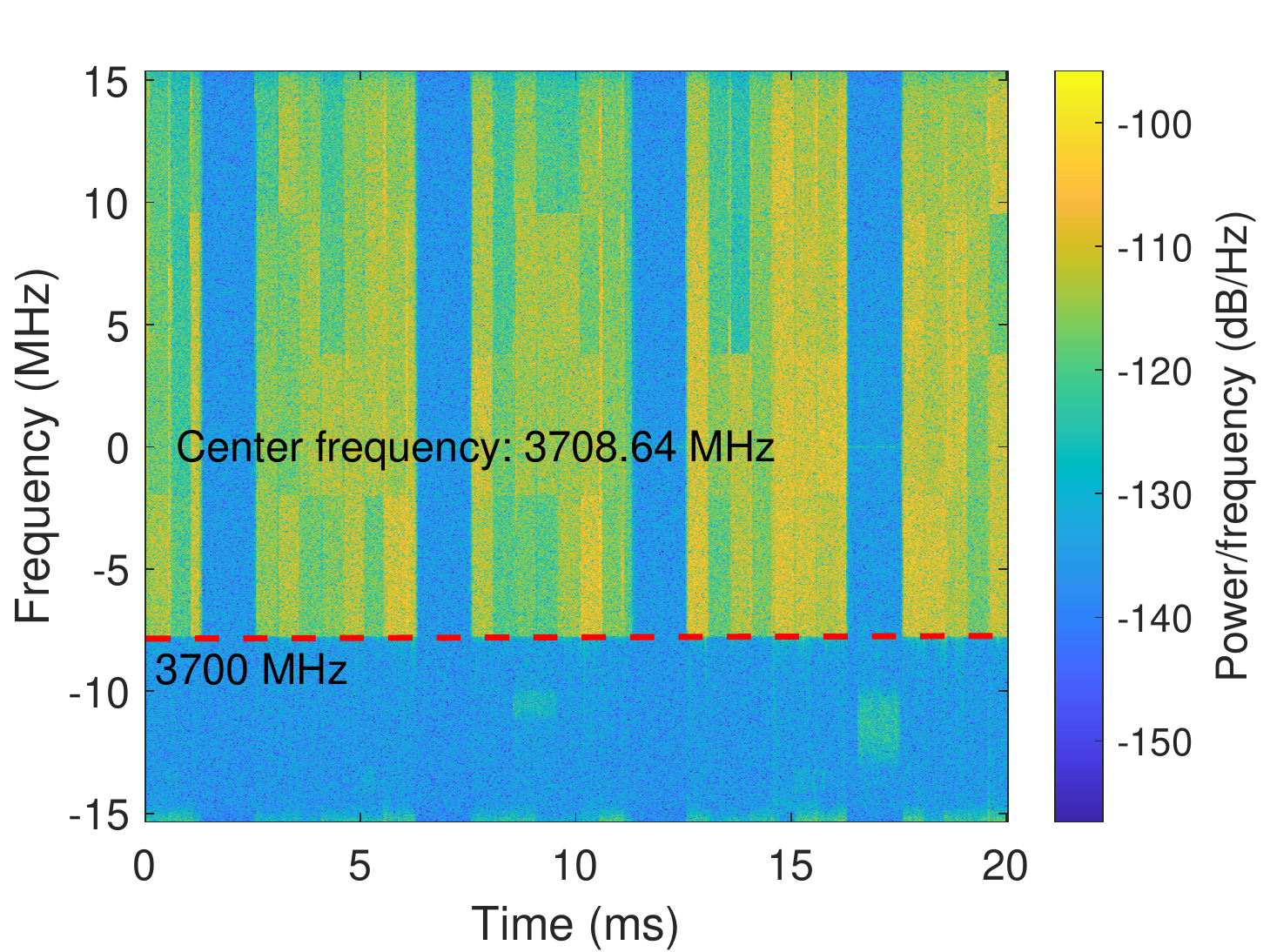}}
        \subfloat[Center frequency: 3730~MHz. Downsampled to 256-FFT with 30~KHz subcarrier spacing to capture bandwidth for a SSB.]{\includegraphics[width=0.48\textwidth]{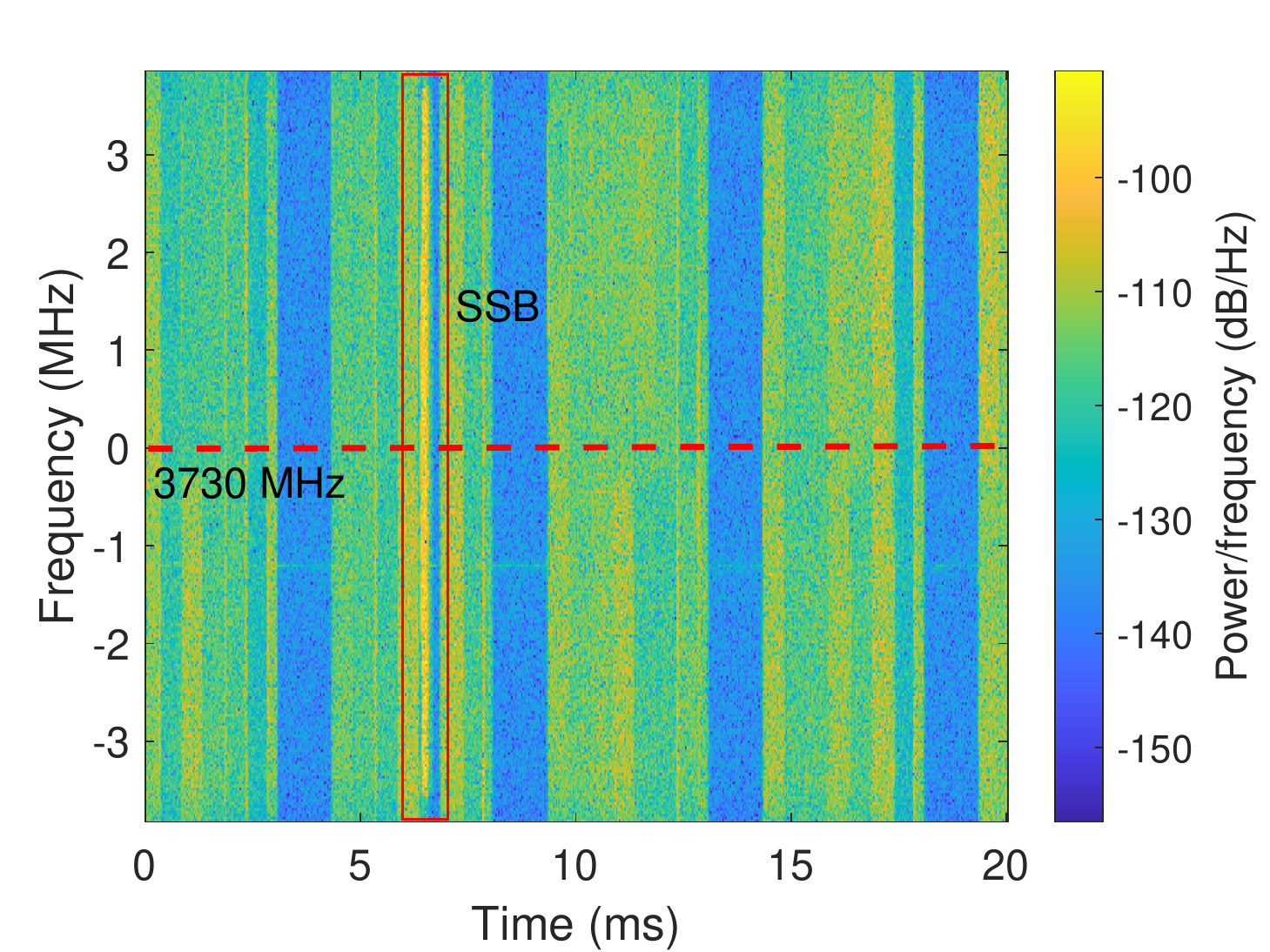}\label{fig:spectro_d}}
	\caption{Spectrogram of measured I/Q samples of 5G NR n77 band at 12:30:30. Three different center frequency spectrums in (a), (b), and (c) show the occupancy of a 60~MHz bandwidth signal.}\label{fig:spectro}
\end{figure*}

In this work, we present measurement and post-processing results of 5G NR raw IQ samples by a helikite in an urban area. The helikite collects public wireless signals from commercial 5G NR BSs in the C-Band (3.7 GHz). We describe the measurement setup and present the 5G NR synchronization process, as well as the reference symbol received power (RSRP) and reference symbol received quality (RSRQ) extraction, which are obtained by the collected raw I/Q samples (AERIQ)~\cite{maeng2022aeriq}. Based on the measurement results, we elaborate on the cyber security and privacy aspects where 5G NR may be vulnerable, especially when it is possible to collect wide-scale surveillance data from aerial platforms.

\section{5G NR I/Q Collection from a Helikite}
In this section, we describe how we collect I/Q datasets using a helikite at NC State University.

\subsection{Experiment Scenario and Setup}
The experiment is conducted at the NSF AERPAW platform~\cite{marojevic2020advanced} at the main campus of NC State University. We float the AERPAW helikite to an altitude of 400 feet throughout the day from noon to 9 p.m. during NC State’s Packapalooza festival in August 2022. A photo taken in the experiment is shown in Fig.~\ref{fig:photo_packa}. The AERPAW helikite is equipped with an SDR (USRP B205mini) and a GPS receiver while collecting I/Q samples. The helikite obtained spectrum sweeps up to 6 GHz during the whole flight for collecting spectrum occupancy data. Moreover, the helikite is set to collect 20~ms I/Q samples every 9 minutes with a 30.72~MHz sampling frequency at the 3.7 GHz C-Band. From the analysis results of spectrum monitoring, we found that 60~MHz bandwidth 5G NR n77 band signals from 3.70~GHz to 3.76~GHz are detected during the experiment, which is operated by Verizon in United States. 

In this paper, we focus on presenting results from the n77 band of 5G NR I/Q measurement and post-processing. The study of AERPAW helikite spectrum occupancy monitoring results can be found in~\cite{maeng2023spectrum,raouf2023spectrum}. The altitude changes during the experiment recorded by the helikite-mounted GPS receiver are shown in Fig.~\ref{fig:altitude}. In addition, the 3D trajectory of the helikite from GPS logs can be found in \cite{maeng2023spectrum}.

\begin{figure*}[t!]
	\centering
	\subfloat[Correlation between a received signal and PSS candidates to estimate $N_{\rm ID}^2$ and coarse frequency offset.]{\includegraphics[width=0.48\textwidth]{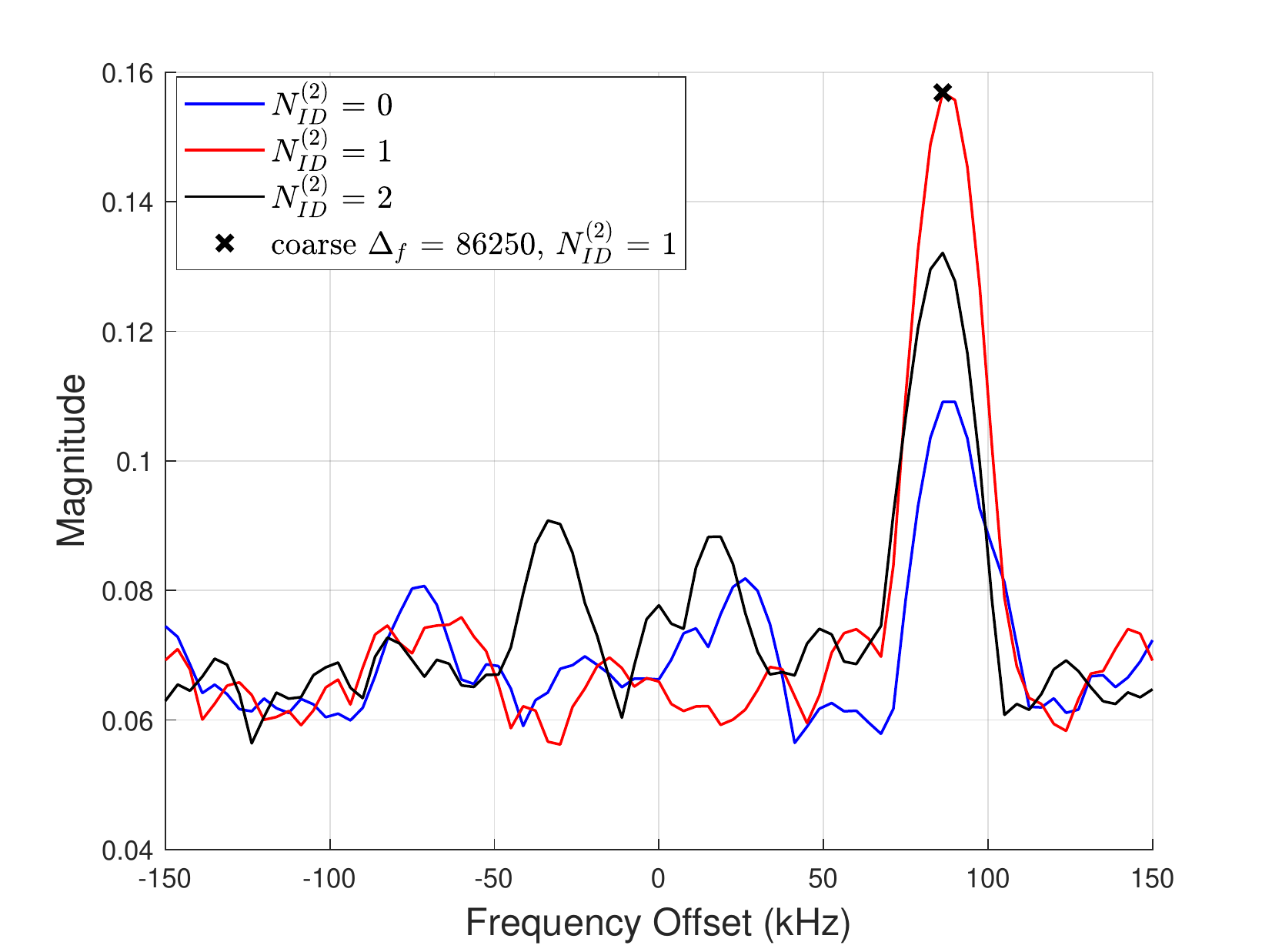}\label{fig:PSS}}
         \subfloat[Correlation between a received SSS and SSS candidates when $N_{\rm ID}^2=1$ is chosen.]{\includegraphics[width=0.48\textwidth]{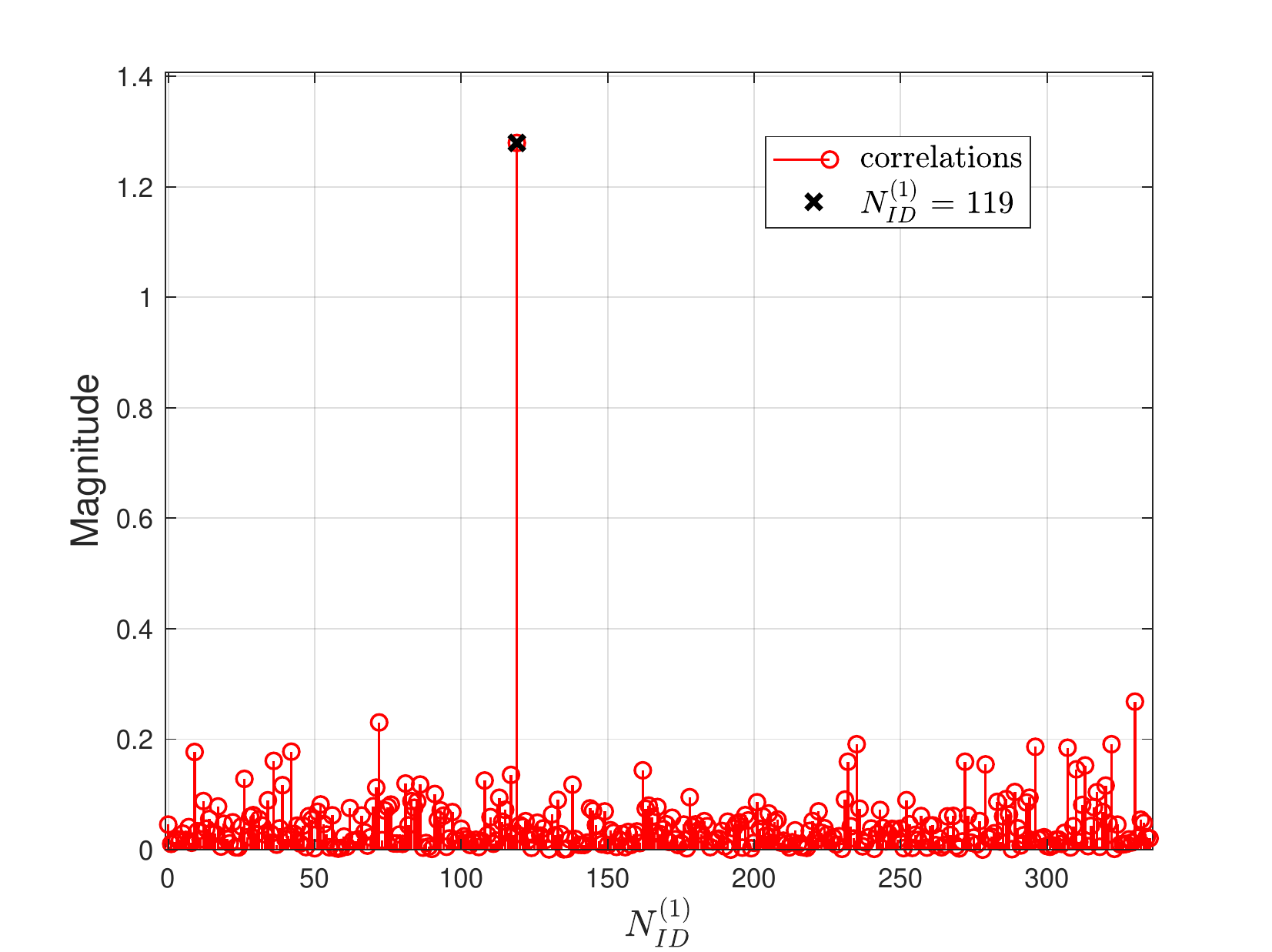}\label{fig:SSS_1}}\\
        
        \subfloat[Correlation between a received SSS and SSS candidates when $N_{\rm ID}^2=2$ is manually chosen.]{\includegraphics[width=0.48\textwidth]{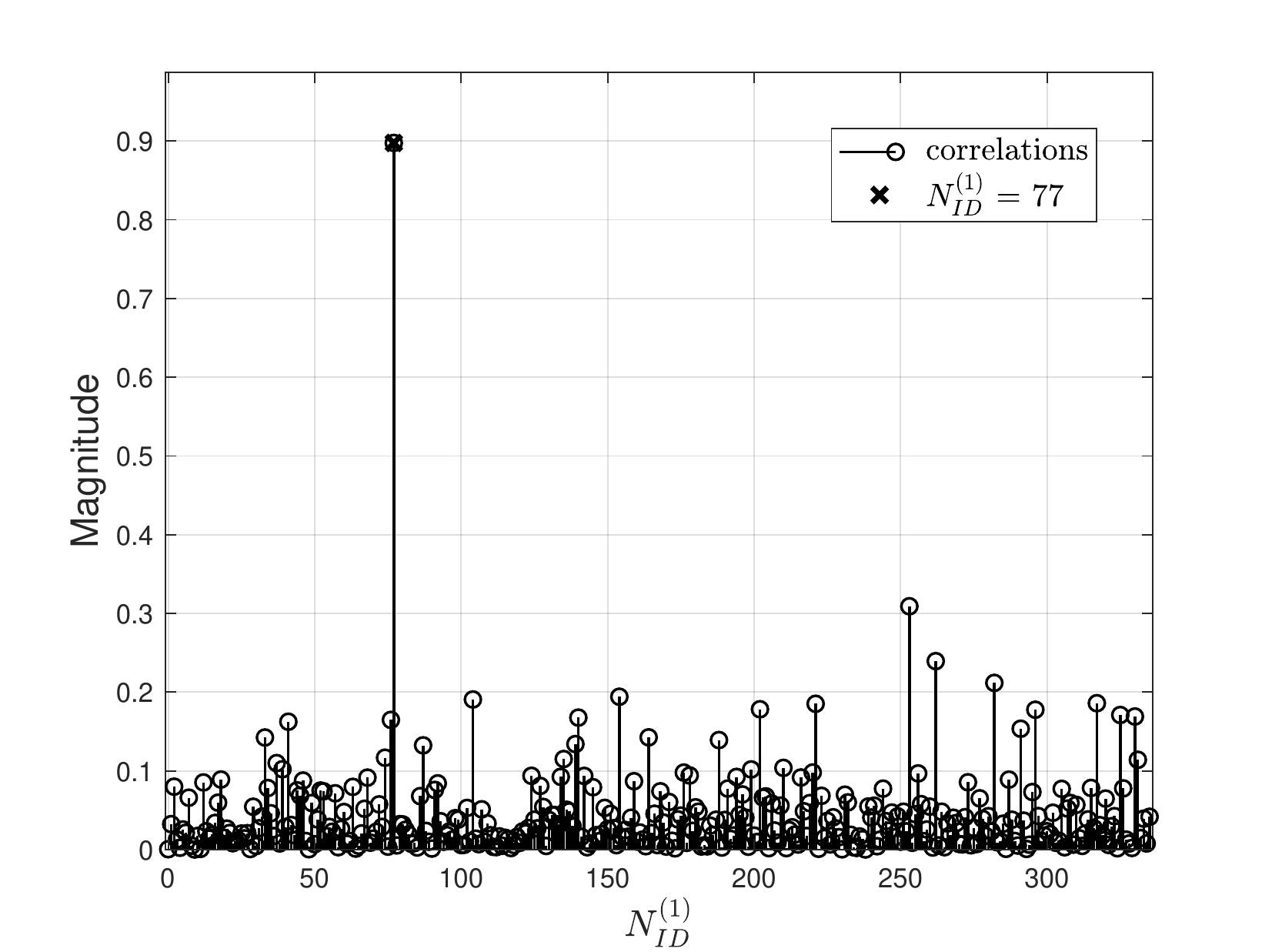}\label{fig:SSS_2}}
        \subfloat[Correlation between a received SSS and SSS candidates when $N_{\rm ID}^2=0$ is manually chosen.]{\includegraphics[width=0.48\textwidth]{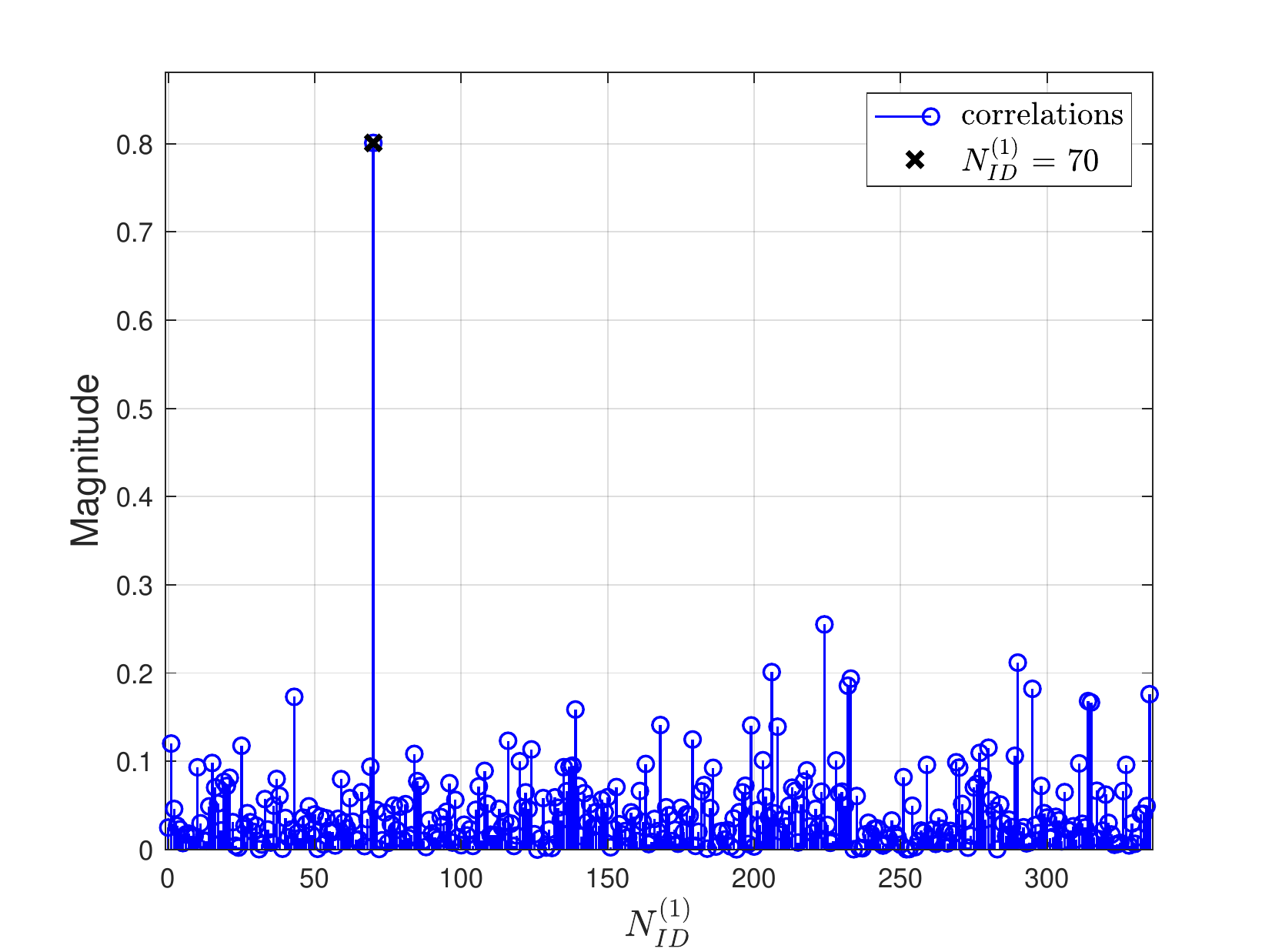}\label{fig:SSS_3}}
	\caption{Correlation peaks to estimate PCI from I/Q samples at 12:30:30. The PSS peak is the highest when $N_{\rm ID}^2=1$ while peaks from other $N_{\rm ID}^2$ are still observed in (a). Three different PCIs are detected in cell search. }\vspace{-0.2in}\label{fig:PSS_SSS}
\end{figure*}

\section{5G NR I/Q Post-processing}
In this section, we present the post-processing of collected 5G NR I/Q samples from the helikite experiment. We utilize MATLAB 5G Toolbox~\cite{5G_Toolbox} to decode I/Q sample datasets.

\subsection{Spectrogram Analysis}
We analyze spectrum occupancy of the 5G NR n77 band from collected I/Q samples by the helikite. Fig.~\ref{fig:spectro} shows spectrograms of I/Q samples for different center frequencies at 12:30:30 (HH:MM:SS), which is a visual representation of the spectrum of frequencies of a signal as it varies over time. The bandwidth of the spectrum is 30.72~MHz following the sampling frequency. The center frequency is not directly reflected in the spectrogram, and it is represented by 0~MHz. We found that an SSB is allocated at 3730~MHz every 20~ms. We tune the center frequency to 3730~MHz and downsample I/Q to 256-FFT with a subcarrier spacing (SC) 30~KHz to adjust the bandwidth to the SSB in Fig.~\ref{fig:spectro_d}.

\begin{figure*}[t!]
	\centering
	\subfloat[RSRP versus time]{\includegraphics[width=0.5\textwidth]{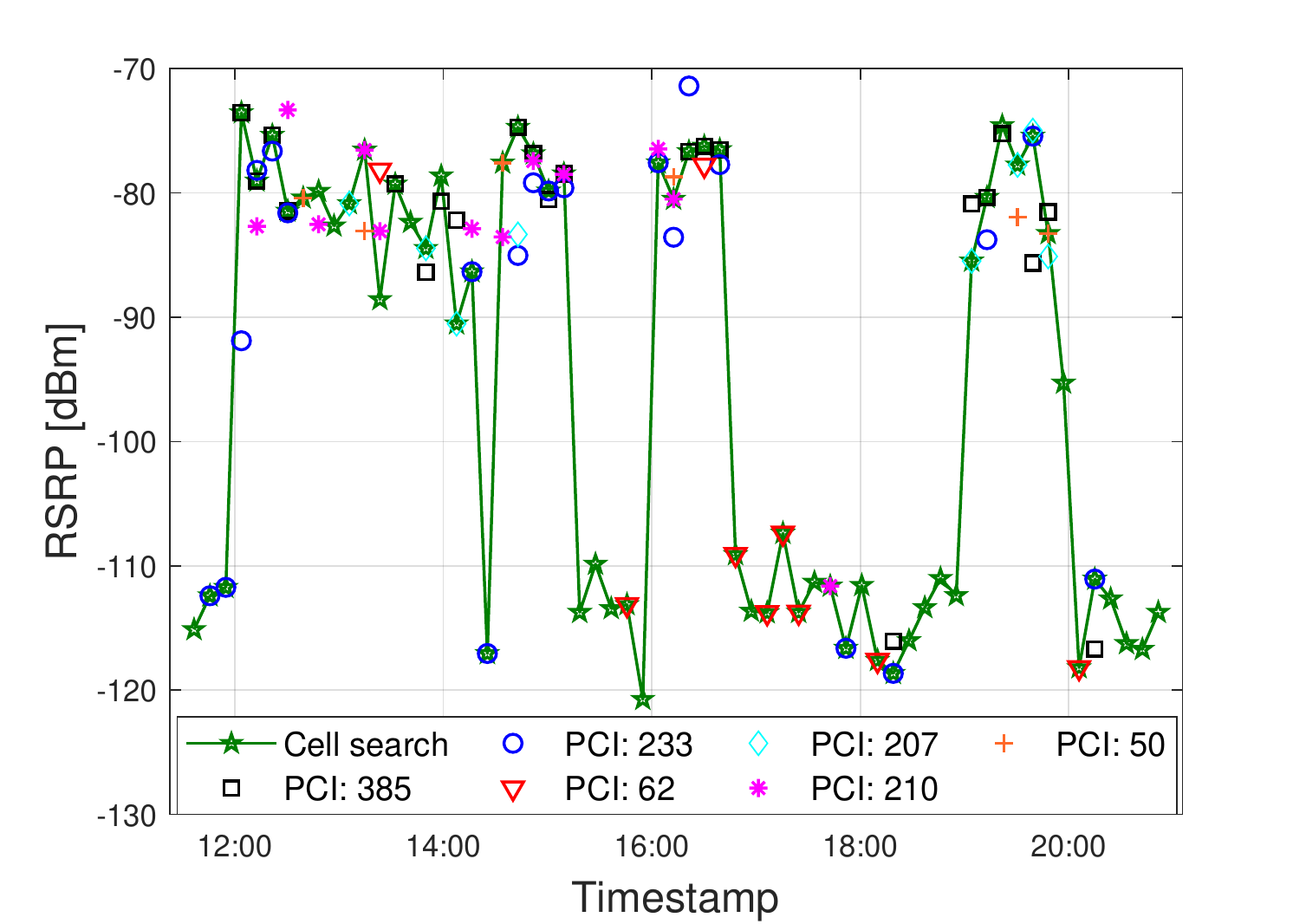}\label{fig:RSRP_t}}
        \subfloat[RSRQ versus time]{\includegraphics[width=0.5\textwidth]{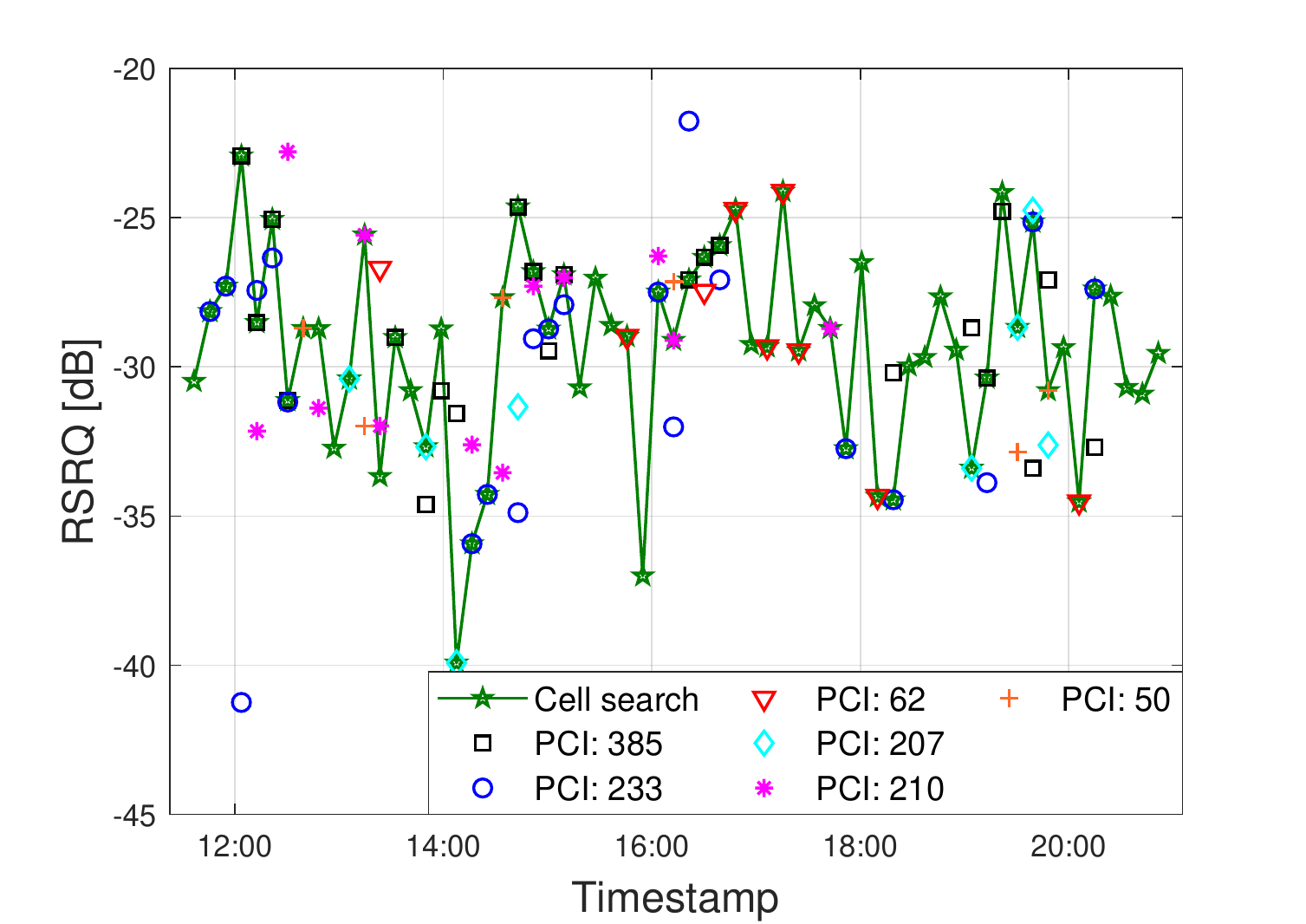}\label{fig:RSRQ_t}}\\
        
        \subfloat[RSRP versus altitude]{\includegraphics[width=0.5\textwidth]{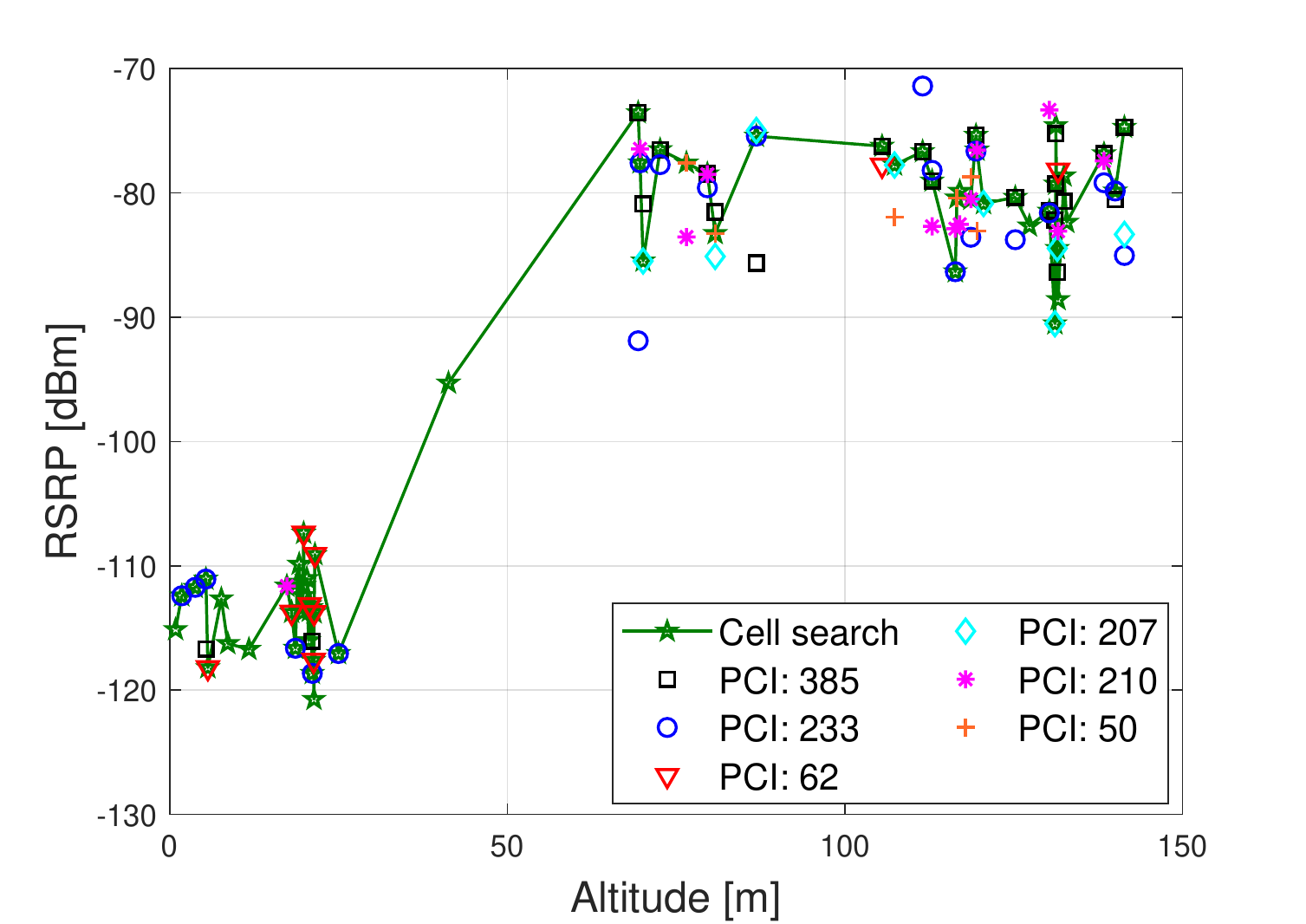}\label{fig:RSRP_alt}}
        \subfloat[RSRQ versus altitude]{\includegraphics[width=0.5\textwidth]{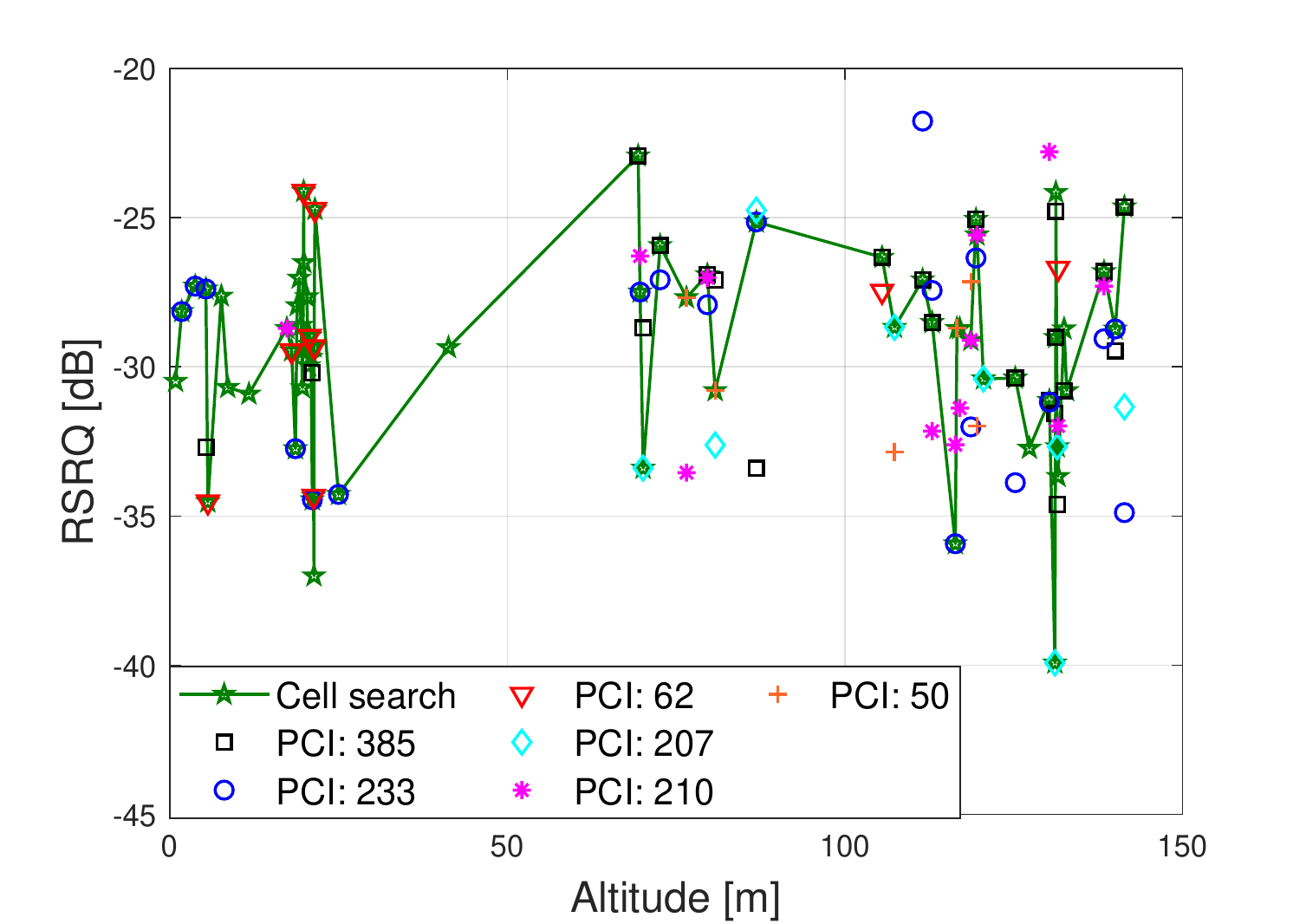}\label{fig:RSRQ_alt}}
	\caption{RSRP and RSRQ are calculated by SSS and PBCH-DMRS in an SSB. Cell search indicates PCIs are estimated by the algorithm in Section \ref{sec:cell_search}, while an algorithm in specific PCI detects the target PCI by a correlation peak analysis and calculates RSRP and RSRQ if the PCI is detected.}\label{fig:RSRP_RSRQ}
\end{figure*}

\subsection{Cell Searach and Synchronization by SSB}\label{sec:cell_search}
In the first stage of 5G NR synchronization, user equipment (UE) compensates frequency offset, estimates $N_{\rm ID}^2$ for physical cell ID (PCI), and estimates timing offset by using a primary synchronization signal (PSS) in the SSB, which is generated by an $m$-sequence (c.f. the Zadoff-Chu sequence in LTE systems). After we downsample I/Q samples to capture SSB as in Fig.~\ref{fig:spectro_d}, we calculate all possible combinations of cross-correlations between the received signal by compensating candidate coarse frequency offset and candidate PSS sequences. A UE chooses the combination of coarse frequency offset and PSS that achieves the highest correlation. Next, the fine frequency offset is compensated by using the cyclic prefix (CP) of orthogonal frequency division multiplexing (OFDM) symbols correlation across the SSB. In the end, the timing offset is estimated by a correlation peak between received signals and the estimated PSS.

In the second stage, user equipment (UE) estimates $N_{\rm ID}^1$ using secondary synchronization signal (SSS) in an SSB, which is generated by a Gold sequence. Then, UE estimates PCI among 1008 candidates by using estimated $N_{\rm ID}^1$ and $N_{\rm ID}^2$ from the formula: $\mathsf{PCI}=3 N_{\rm ID}^1 + N_{\rm ID}^2$. The highest correlation between received SSS and all candidates of SSS given $N_{\rm ID}^2$ is selected to estimate $N_{\rm ID}^1$.

Fig.~\ref{fig:PSS_SSS} shows correlation peaks while estimating PCI by the PSS and the SSS from collected I/Q samples at 12:30:30 (HH:MM:SS). In Fig.~\ref{fig:PSS}, the highest peak is observed at  $N_{\rm ID}^2=1$ and the estimated coarse frequency offset is 86.25~KHz. The corresponding $N_{\rm ID}^1$ estimation by correlation peak of the SSS  ($N_{\rm ID}^2=1$) is shown in Fig.~\ref{fig:SSS_1}. However, we can also observe correlation peaks of a PSS from other $N_{\rm ID}^2$ and we detect different PCIs by manually choosing $N_{\rm ID}^2=2, 0$ in Fig~\ref{fig:SSS_2} and Fig~\ref{fig:SSS_3}. This implies that the SDR at the helikite has received signals from three different base stations (BSs). By analyzing cell search results during the whole experiment, we conclude that at least 6 different BSs signals are detected from the helikite during the whole data collection period.

\subsection{RSRP and RSRQ by SSB}
In 5G NR, RSRP (SS-RSRP) and RSRQ (SS-RSRQ) can be calculated by SSS and the PBCH-DMRS in an SSB, which is used to measure received signals power and quality of signals from BSs. Since sequences of PBCH-DMRS and SSS are unique in terms of a PCI, RSRP and RSRQ can be calculated by separate PCIs. Fig.~\ref{fig:RSRP_RSRQ} shows RSRP and RSRQ depending on the time and altitude of the helikite. We obtain RSRP and RSRQ by the cell search algorithm. The individual 6 dominant PCIs during the experiment are also obtained by cell search results. In particular, the cell search algorithm indicates that we estimate a single PCI by using the cell search algorithm in Section~\ref{sec:cell_search} and calculate RSRP and RSRQ corresponding to the PCI. 

For a specific PCI, we first manually detect the PCI using PSS and SSS correlation peaks and calculate RSRP and RSRQ for that specific PCI. We drop the calculation of RSRP and RSRQ if the PCI is not detected by analyzing the correlation peak of PSS and SSS. Therefore, not all of the RSRP and RSRQ for specific PCIs are marked in Fig.~\ref{fig:RSRP_RSRQ}. In Fig.~\ref{fig:RSRP_alt}, we observe that RSRP increases as the altitude of the helikite increases until a certain altitude and after the altitude, RSRP does not increase. On the other hand, the dependency of altitude is not clear in RSRQ in Fig.~\ref{fig:RSRQ_alt}. In addition, RSRQ is low regardless of the level of RSRP, which implies that interference from multiple BSs is high during the whole experiment.

\section{Discussion: Security and Privacy Considerations for 5G}

In this section, we elaborate on various different aspects where 5G systems may be vulnerable in terms of surveillance, security, and privacy.   

\subsection{Surveillance}

Our results in this paper show that using a commercially available SDR, it is possible to capture signals from a 5G network. In particular, it is possible to extract RSRP and RSRQ signals after post-processing the raw I/Q samples. For these collected I/Q measurements we have had difficulty in extracting the data in the physical broadcast channel (PBCH) as the propagation channel was very frequency selective. However, with improvements in the data collection process, and improvements to the RF front end of the USRP (no RF front end to the USRP was used for the results reported in this paper), it would be possible to decode additional channels such as the PBCH, the physical downlink control channel (PDCCH), among others, and to improve the coverage range. We have collected very sparse I/Q data as our goal was to analyze how the RSRP and RSRQ changed with altitude, and it would be possible to capture data more densely over time.  

The ease of capturing I/Q samples easily with commercial SDR equipment has implications for security and privacy vulnerabilities of existing and future 5G/NextG wireless networks. As was recently seen in the surveillance balloon incident that flew across the United States~\cite{oxford2023balloon}, it is possible to collect large volumes of data at critical frequencies and from critical infrastructure. Flying at higher altitudes makes it possible to have line-of-sight with more transmitters, hence it becomes possible to capture signals from virtually any signal source as long as the received signal strength is strong enough~\cite{raouf2023spectrum}. High-end SDRs rather than the USRP B205mini can be used to improve the signal reception quality and hence the coverage range. This may introduce a security concern especially for critical infrastructure and bands. For example, in the CBRS band, storing information about a radar transmitter's movement and position is not allowed by regulations~\cite{sarkar2021deepradar}. There may also be other legal implications and consequences of recording and storing raw spectrum data in certain bands; for a related discussion, the readers can see~\cite{sicker2007legal}.

\subsection{Jamming Resilience}

Various ways that 5G NR may be vulnerable to jamming, spoofing, and sniffing have been studied in~\cite{lichtman20185g,arjoune2020smart,girke2019towards}. 
In particular, synchronization signals (SS), physical broadcast channel (PBCH), and the physical broadcast channel demodulation reference signal symbols (PBCH-DMRS) in a synchronization signal block (SSB) are easy targets from an adversary in terms of attack efficiency and complexity~\cite{lichtman20185g}. When compared to LTE, 5G NR may improve in some ways the resilience to jamming attacks. For instance, LTE Physical Control Format Indicator Channel (PCFICH) has been removed in 5G NR, and PUCCH, PSS, and SSS are allocated in more dynamic locations in a resource grid.


Most cellular systems implement carrier aggregation, i.e. aggregate a number of channels across many different bands. For example, based on our past measurements of deployed cellular networks~\cite{sathya2020measurement,rochman2021comparison,infocom2022}, we see that 4G systems routinely aggregate across multiple licensed bands as well as up to 3 unlicensed channels in 5 GHz (20 MHz each) and shared bands such as CBRS. Carrier aggregation in 5G today is mostly seen in millimeter wave (mmWave) where up to eight 100 MHz channels are aggregated. Carrier aggregation in the mid-bands is not yet widely implemented, but 3GPP allows up to 16 channels to be aggregated. Aggregation allows a certain level of resiliency to jamming since an adversary would have to jam all possible channels that a BS might use, including unlicensed and shared bands and would have to intercept many channels across different bands to decode information being transmitted since the primary channel and the channels being aggregated can change on a ms level. If a small subset of these channels is jammed, while there may be a slight loss in capacity, the overall network can continue to operate on the other available channels. Channel aggregation is also permitted across FR1 and FR2 bands leading to further resiliency. 

\subsection{Post-Quantum Era Attacks}

As discussed earlier in this section, it may be possible to capture and store large volumes of wireless I/Q data at high-altitude platforms by malicious entities. The user-plane data are encrypted for the existing cellular networks, and hence, it is presently not possible to decrypt them with today's computing capabilities from raw I/Q samples. However, quantum computing capabilities in the post-quantum era are expected to be powerful enough to break through today's encryption capabilities. There are several works in the literature on   
harvest now decrypt later (HNDL) or store now decrypt later (SNDL) attacks~\cite{garcia2022disruptive,raddo2019quantum}. Hence, it carries critical importance to start adopting post-quantum  public-key cryptography approaches that are not vulnerable to HNDL/SNDL attacks~\cite{alagic2020status}.

\section{Concluding Remarks}

In this paper, we collect 5G NR I/Q sample datasets in the n77 band (3.7 GHz C-band) by using a helikite in an urban area. The helikite is equipped with an SDR and a GPS receiver. We decode 5G NR I/Q samples by MATLAB 5G Toolbox and present a spectrogram for spectrum occupancy, cell search and synchronization using PSS and SSS in an SSB, and RSRP and RSRQ by SSS and PBCH-DMRS in an SSB. Correlation peaks from PSS and SSS are observed, and the RSRP and RSRQ are shown with respect to the altitude of the helikite. We observe that the RSRP increases as altitude increases, while the dependence of RSRQ to altitude is not as strong. In addition, we discuss the security and privacy aspect of 5G NR including vulnerabilities to surveillance, jamming, and post-quantum era attacks.

\bibliographystyle{IEEEtran}
\bibliography{IEEEabrv,references}
  
\end{document}